\documentclass[useAMS]{mn2e}
\usepackage{graphicx}
\usepackage{latexsym}
\usepackage {keyval}
\usepackage{amsmath}
\title[Orbital release model]
{An orbital release model for the Orion BN/KL fingers}
\author[Raga, A. C., Rivera-Ortiz, P. R.,
  Cant\'o, J.,  Rodr\'\i guez-Gonz\'alez, A.,
  Castellanos-Ram\'\i rez, A.]{
  A. C. Raga$^{1}$\thanks{E-mail: raga@nucleares.unam.mx},
  P. R. Rivera-Ortiz$^2$,
  J. Cant\'o$^3$,
  A. Rodr\'\i guez-Gonz\'alez$^1$ \and
  A. Castellanos-Ram\'\i rez$^3$\\
$^1$Universidad Nacional Aut\'onoma de M\'exico,
Instituto de Ciencias Nucleares,
AP 70-543, CDMX 04510, M\'exico\\
$^2$ Observatoire de Grenoble, Grenoble Alpes, CNRS, IPAG,
38000 Grenoble, France\\
$^3$ Universidad Nacional Aut\'onoma de M\'exico,
Instituto de Astronom\'\i a,
AP 70-264, CDMX 04510, M\'exico\\
}
\begin{document}

\date{}

\pagerange{\pageref{firstpage}--\pageref{lastpage}} \pubyear{2019}

\maketitle

\label{firstpage}

\begin{abstract}
  We present a simple model in which the bullets that produce
  the ``Orion fingers'' (ejected by the BN/KL object) are interpreted
  as protoplanets or low mass protostars in orbit around a high
  mass star that has a supernova explosion. As the remnant of the
  SN explosion has only a small fraction of the mass of the pre-supernova
  star, the orbiting objects then move away in free trajectories, preserving
  their orbital velocity at the time of release. We show that a system
  of objects arranged in approximately co-planar orbits results in trajectories
  with morphological and kinematical characteristics resembling the
  Orion fingers. We show that, under the assumption of constant
  velocity motions, the positions of the observed heads
  of the fingers can be used to reconstruct the properties of the
  orbital structure from which they originated, resulting in a compact
  disk with an outer radius of $\sim 2.4$~AU.
\end{abstract}


\section{Introduction}

One of the more intriguing flows in the Galactic ISM is composed
of the ``fingers'' radiating away from the Orion BN/KL object
(a massive embedded stellar object surrounded by a compact nebula,
see Becklin \& Neugebauer 1967 and Kleinmann \& Low 1967).
These fingers were first reported by Allen \& Burton (1993)
in IR H$_2$ and [Fe~II] emission, and have lengths of up to
$\sim 2'$ ($\sim 7\times 10^{17}$~cm at a distance of 400~pc).
The bases of these fingers, as well as a multitude of other
filaments are detected in interferometric CO observations
(Zapata et al. 2009; Zapata et al. 2017; Bally et al. 2017).
The CO emission of the individual fingers shows a remarkable
``Hubble law'' of linearly increasing radial velocities with
distance from the BN/KL object.

Doi et al. (2002) and Bally et al. (2011, 2015) obtained
proper motions of the IR line emission of the heads
of the ``Orion fingers'', and Nissen et al. (2007)
obtained radial velocities of these features. Nissen
et al. (2012) combined proper motion and radial velocity
measurements to obtain the full spatial motion of some
of the fingers.

The fact that the motions of the heads of the individual fingers
give similar dynamical timescales implies that
they were all ejected in an ``explosive'' event of short duration.
This result stands regardless of the fact that some of the ``bullets''
at the tips of the fingers might have had a substantial slowing
down along their trajectories (Rivera-Ortiz et al. 2019a, b).

It has been proposed that this sudden event could have been:
\begin{itemize}
\item a result of the breakup of a multiple stellar system
  (see Reipurth et al. 2010; Reipurth \& Mikkola 2012, 2015;
  Moeckel \& Goddi 2012). The dynamics of a 3- or 4-body system
  can explain the high velocities of the stars near the BN/KL
  object (observed in radio continuum by Rodr\'\i guez et al.
  2005 and by G\'omez et al. 2005, 2008),
  and in principle could trigger
  the ``interstellar bullets” that create the Orion fingers.
  A first attempt to model the ejection of a cluster of bullets
  through a gravitational interaction was recently presented
  by Rivera-Ortiz et al. (2021),
\item a result of a supernova, which could liberate 
  the bound orbital motion of companion stars into free
  hyperbolic orbits. This scenario can
  explain the high velocities of the stars around BN/KL, and
  could as well result in the ejection of the observed
  high velocity bullets.
\end{itemize}
These scenarios are discussed in the previous literature on
the Orion fingers (see, e.g., Bally et al. 2011, 2017
and Zapata et al. 2017).

A remarkable feature of the outflow from BN/KL is that
while the whole system of fingers shows an elongated
morphology towards the NW and SE, this roughly bipolar
structure does not correspond to a division between
blue and redshifted emission. There is a ``red/blueshift''
division in the system of fingers, but with respect
to an axis roughly aligned with the NW/SE elongation.
This NE/SW division of the blue (NE) and redshifted (SW)
emission is clearly seen in Figure 3 of Zapata et al. (2017).
However, there are some filaments with the
``wrong radial velocity shift'' (i.e., some redshifted
filaments to the NE and some blueshifted filaments
to the SW), as can be seen in the observations
of Bally et al. (2017).

In the present paper we show that the elongated morphology
and ``side-to-side'' red/blue shifted kinematics of the
system of Orion fingers can be straightforwardly explained
as the free propagation of a system of objects with a planar
initial velocity distribution. In section 2 we show how
this extremely simple model produces spatial and radial
velocity distributions that qualitatively resemble the
system of Orion fingers. Section 3 derives the relation
between the radial (position) distribution function and the orbital
velocity distribution of a system of massless particles in circular
orbits around a massive object. In section 4 we show how under
the assumption of a planar velocity distribution and free
post-supernova trajectories the
observed positions (and also velocity information if
available) can be used to reconstruct the velocity
distribution of the bullet system. In section 5
we use the observed positions of the tips of the Orion fingers
to reconstruct their velocity distribution and the radial
position distribution of the bullets on the pre-release
circumstellar disk. The results are discussed in section 6.

\section{Instantaneous release model}

\subsection{General considerations}

We assume that we have a system of $N$ non-interacting particles
with initial (at $t=0$) positions $\underline{x}_{0,i}$ and velocities
$\underline{v}_i$ (with $i=1,\, ...\,N$). For $t>0$, the free
particles then have (vector) positions:
\begin{equation}
  \underline{x}_i(t)=\underline{x}_{0,i}+t\,\underline{v}_i\,.
  \label{xi}
\end{equation}
If initially the particles lie within a volume of size $L$,
it is clear that for $t\gg L/v_{m}$ (where $v_{m}$ is
the smallest modulus of all of the particle velocities) the first
term on the right hand side of equation (\ref{xi}) can be neglected,
and we have trajectories that are independent of the
initial positions of the particles, and only depend on their initial
velocities.

We present two examples of this trivial model, a ``thin disk''
(S 2.2) and a ``thick disk'' initial distributions (\S 2.3). The
thin disk distribution has a system of particles in coplanar
circular orbits around a central star, and the thick disk distribution
has circular orbits with a range of inclinations of their orbital axes.
At $t=0$ the central star has a supernova explosion, ejecting most of its
mass: a remnant of only $\sim 2$\,M$_\odot$ will be left behind when a
$\sim 30$\,M$_\odot$ solar metallicity star has a SN explosion (see. e.g.,
Liu et al. 2021 and references therein). With such a drastic decrease
in the mass of the central star, the orbiting particles will suddenly
find themselves in free, very open hyperbolic orbits, which can be approximated
as linear trajectories.

The question then is what are these particles in initial circular
orbits around the pre-supernova, massive star: obvious candidates
are giant planets/protoplanets and/or very low mass stars forming
in a disk around the massive star. This gives the attractive picture
of the Orion fingers being produced by ejected pre-planetary or pre-stellar
condensations.

\subsection{Planar distribution with circular orbits}

We now consider a planar distribution of 100 massless particles
orbiting around a central star of mass $M_*$, with orbital
radii $r$ distributed randomly in the $0\to R_0$ range. The phase
of the orbit at $t=0$ is chosen uniformly in the $0\to 2\pi$
range. Their orbital velocities then are:
\begin{equation}
  v_{orb}=v_0\sqrt{\frac{R_0}{r}}\,,
  \label{vorb}
\end{equation}
with
\begin{equation}
  v_0\equiv \sqrt{\frac{GM_*}{R_0}}\,.
  \label{v0}
\end{equation}
The combination of $v_0$ and $R_0$ gives a characteristic time
\begin{equation}
  t_0\equiv \frac{R_0}{v_0}=\sqrt{\frac{R_0^3}{GM_*}}\,.
  \label{t0}
\end{equation}

For example, for a distribution of particles
with an outer radius $R_0=100$~~AU orbiting around
a central star with a $M_*=30M_\odot$ mass, we have
$v_0=16.3$~km~s$^{-1}$ and $t_0=29.1$~yr.

Assuming that at $t=0$ the central star suddenly ``disappears'',
the particles then evolve with straight trajectories, preserving
their initial velocity. Figure 1 shows the particle trajectories
up to a time $300\,t_0$, projected onto the plane of the sky for
four different assumed angles between the disk axis and the line of
sight.


\begin{figure}{\centering}
\includegraphics[width=\columnwidth]{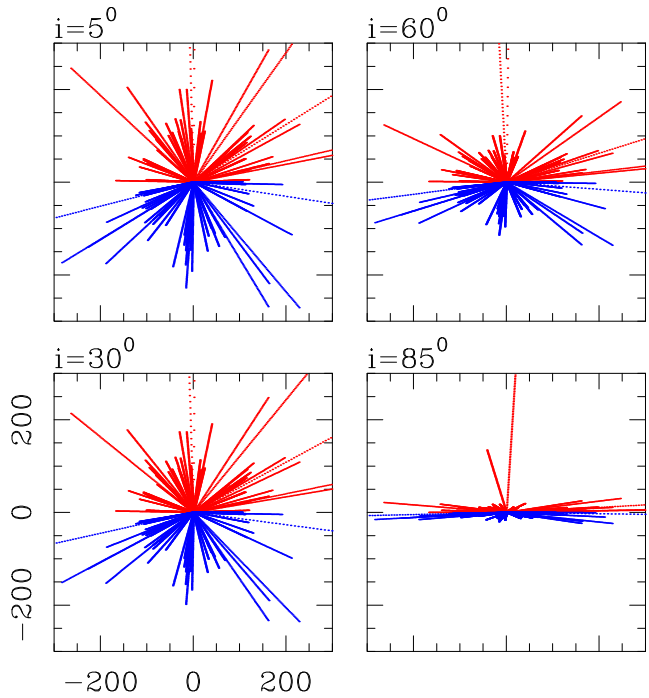}
\caption{Trajectories of a system of free particles originating
  from a thin disk of outer radius $R_0$ and outer orbital velocity $v_0$
  released at $t=0$ and ending at $300\,t_0$ (see the text). The four
  frames show the projected trajectories for 4 different values of the
  inclination of the initial orbital axis with respect to the line
  of sight. The colours of the trajectories correspond to positive (red)
  and negative (blue) radial velocities of the ejected particles.
  The $(x',y')$ plane of the sky axes are labelled in units of $R_0$.}
\label{fig1}
\end{figure}

\subsection{Non-planar distribution with circular orbits}

We now consider the same problem of 100 massless particles
in initial circular orbits (which are ``liberated'' at $t=0$, see \S 2.2),
but allowing the individual orbits to have axes with random
tilts in the $0\to 35^\circ$ range with respect to the average
disk axis.

The resulting particle trajectories at $t=300\,t_0$ (see equation
\ref{t0}) are shown in Figure 2. We see that the variation in the
initial orbital axes of the particles results in a partial mixing
of redshifted and blueshifted particles above/below the $y'=0$
axis. This effect is more strongly pronounced at the smaller
and larger values of the inclination ($i=5$ and $85^\circ$, see
Figure 2).
  
\begin{figure}{\centering}
\includegraphics[width=\columnwidth]{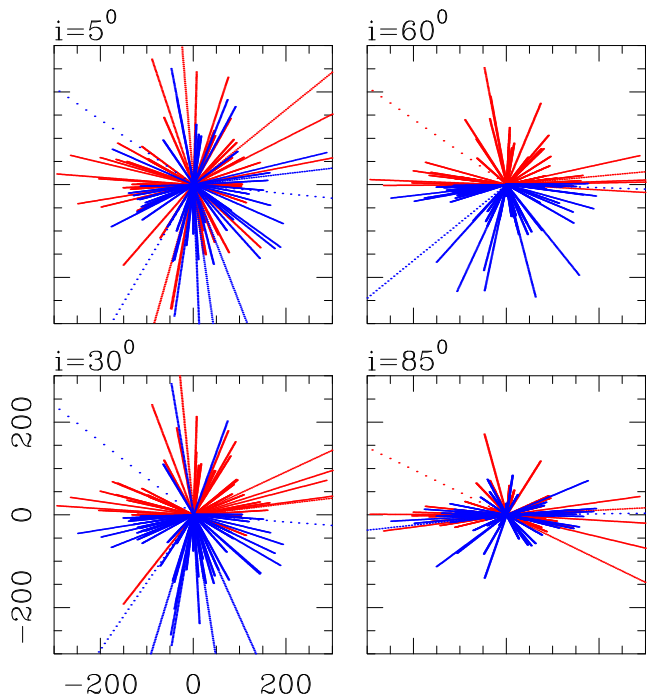}
\caption{The same as Figure 1 but for an initial ``thick disk''
  distribution of orbits with axes with random tilts of up to $35^\circ$
  with respect to the disk axis.}
\label{fig2}
\end{figure}

\section{The relation between the radial density and velocity
  distributions of particles in an orbital disk}

In \S 2.2 we considered
a set of $N$ particles in co-planar circular orbits around a
central source, with a uniform radial distribution. Let us now
look at a more general problem, in which the radial particle
distribution has an arbitrary power law form:
\begin{equation}
  f(r)=\frac{(\alpha+1)N}{R_0^{\alpha+1}}r^\alpha\,,
  \label{fr}
\end{equation}
where the proportionality constant is determined from
the $\int_0^{R_0} f(r)dr=N$ condition, with $r$ being the radius on
the orbital plane and $R_0$ being the radius of
the outer orbit of the $N$ particles.

It is straightforward to show that this distribution corresponds
to a surface number density of particles:
\begin{equation}
  n(r)=\frac{(\alpha+1)N}{2\pi R_0^{\alpha+1}}r^{\alpha-1}\,.
    \label{nr}
\end{equation}

As the particles are in circular orbits, their velocity
as a function of $r$ is given by equation (\ref{vorb}).
We can then calculate the distribution $g(v)$ of the modulus $v$ of
the velocities of the particles as:
\begin{equation}
  g(v)=\frac{f(v)}{|dv/dr|}=\frac{2(\alpha+1)Nv_0^{2+2\alpha}}{v^{3+2\alpha}}\,,
  \label{gv}
\end{equation}
where equations (\ref{vorb}) and (\ref{fr}) have been used for the
second equality. For the case of a uniform radial particle distribution
(with $\alpha=0$) described in \S 2.2, this gives
\begin{equation}
  g(v)=\frac{2Nv_0^2}{v^3}\,.
  \label{gvu}
\end{equation}

\section{Reconstructing the orbital disk from the evolved trajectories}

\subsection{The ballistic trajetories}

We now consider the ballistic trajectories of particles
at large times, when they have reached distances $L\gg R_0$,
where $R_0$ is the size of the original disk around a star
that at $t=0$ became a SN, thus releasing the particles from
the central, stellar potential. For $t\gg t_0$ (see equations
\ref{xi} and \ref{t0}), the trajectory of a particle is:
\begin{equation}
  x(t)=tv\cos\theta\,;\,\,\, y(t)=tv\sin\theta\,,
  \label{xyt}
\end{equation}
where $\theta$ is the (random) angle between the orbital velocity
and the $x$-axis and $v$ is modulus of the orbital velocity
of one of the particles.

If we assume that the initial disk is at an inclination $i$ with respect
to the line of sight (with the $y$-axis into and the $x$-axis on
the plane of the sky), the plane of the sky trajectory will be:
\begin{equation}
  x'(t)=x(t)\,;\,\,\, y'(t)=y(t)\cos i\,.
  \label{xypt}
\end{equation}
The parcel has plane of the sky velocities
\begin{equation}
  v_{x'}=v\cos\theta\,;\,\,\, v_{y'}=v\sin\theta\cos i\,,
  \label{vxy}
\end{equation}
a radial velocity (with positive values away from the observer)
\begin{equation}
  v_r=v\sin\theta\sin i\,,
  \label{vr}
\end{equation}
and a proper motion velocity
\begin{equation}
  v_{pm}=v\sqrt{\cos^2\theta+\sin^2\theta\cos^2 i}\,.
  \label{vpm}
\end{equation}

Finally. from equation (\ref{vxy}) we compute the average value
over all bullets of the moduli of $x'$ and $y'$:
$$<|x'|>=t<v><|\cos \theta|>\,;$$
\begin{equation}
  <|y'|>=t<v><|\sin \theta|>\cos i\,,
  \label{xyav}
\end{equation}
where we have assumed that $v$ and $\theta$ are uncorrelated. Also,
$<|\cos \theta|>=<|\sin \theta|>=2/\pi$ for random directions
on a plane.

\subsection{The reconstruction}

We assume that at a time $t\gg t_0$ we observe the positions of a system
of particles arising from an orbital disk system that at $t=0$ lost its
central gravitational binding source. We assume that we have
the plane of the sky positions for $N$ observed ``bullets''.
If for these bullets we have observed proper motions and radial
velocities, we can directly calculate the distribution function of
the velocity moduli of the ensemble of particles.

If we have the radial velocities $v_r$ and/or the proper motions
$v_{pm}$ for only some of the bullets (the proper motion of at least
one bullet being necessary), using the assumption that the motions
lie on a plane (i.e., the plane of an initial orbiting disk structure),
we can still reconstruct the velocity distribution function, To
this effect, we proceed with the following steps:

\begin{itemize}
  \item {\it Step 1:}
If the observed system lends itself to an interpretation as an
``unbound debris disk'', we should observe a clear side-to-side
division between positive and negative radial velocities (see Figure
1). The line on which the radial velocities change sign defines
the $\theta=0$ direction, and then the $(x',y')$ coordinates correspond
to the distances along and across this axis (respectively), with
the origin located at the point of divergence of the particle
trajectories.

\item {\it Step 2:}
We then use the bullets for which we have observed proper motions
to calculate the time $t$ since the gravitational release through
the relation $t=\sqrt{x'^2+y'^2}/{v_{pm}}$,
which (apart from being evidently correct) is obtained from equations
(\ref{xyt}), (\ref{xypt}) and (\ref{vpm}). For an instantaneous
release model to be appropriate, similar values of $t$ should be
obtained for all of the bullets with observed proper motions.


\item {\it Step 3:}
  We now calculate the averages
  of the moduli of the plane of the sky $(x', y')$ coordinates
  over all bullets (which we call $<|x'|>$ and $<|y'|$, respectively)
  and use equation (\ref{xyav}) to find the inclination angle as:
  \begin{equation}
    i=\cos^{-1}\left(\frac{<|y'|>}{<|x'|>}\right)\,.
    \label{ip}
  \end{equation}
  
\item {\it Step 4:}
  With the computed values of $t$ and $i$, we now use the relation:
  \begin{equation}
    v=\frac{|x'|}{t}\sqrt{1+\frac{y'^2}{x'^2\cos^2 i}}\,,
    \label{vxyt}
  \end{equation}
  to calculate the full velocity $v$ for all of the individual
  observed bullets (equation \ref{vxyt} can be directly obtained
  from equations \ref{xyt}-\ref{xypt}). These velocities can
  be appropriately binned to obtain the empirical distribution function
  $g_e(v)$ of the velocities of the particles.

\end{itemize}

\section{Reconstruction of the velocity distribution of the Orion fingers}

We took the dataset of Bally et al. (2011, see their Figure 1),
who analyse two epochs of H$_2$ images to obtain the positions and
proper motions of the Orion finger bullets. We then followed the
procedure described in \S 4:
\begin{itemize}
\item {\it Step 1.} From the CO radial velocities of Zapata et al.
  (2009, their Figure 1), we see that the change of sign in the
  radial velocities lies at a $52^\circ$ angle (measured W from N). We then
  calculate the $(x',y')$ coordinates of $N=168$ bullets (with $x'$ along
  the change of sign axis).
\item {\it Step 2.} We can obtain $t$ from the positions and
  proper motion velocities of the bullets with Eq. (15).
  Most of the resulting ages are below 1000 years, and we assumed a
  representative age of 500 yr, equal to the dynamical age of the
  protostellar objects around Orion BN/KL (see Rodr\'iguez et al. 2005).
\item {\it Step 3.} We calculate the average values $<|x'|>$
  and $<|y'|>$ of the moduli of the coordinates, and use equation
  (\ref{ip}) to obtain $i=71^\circ$.
\item {\it Step 4.}  We calculate the moduli of the velocities of
  all particles with equation (\ref{vxyt}), and group them in
  33~km~s$^{-1}$ wide bins to obtain the empirical velocity distribution
  function $g_e(v)$ (normalized so that $\int g_e(v)=N$, where
  $N=168$ is the number of bullets in our sample). The resulting velocity
  distribution function is shown in Figure 3.
\end{itemize}

\begin{figure}{\centering}
\includegraphics[width=\columnwidth]{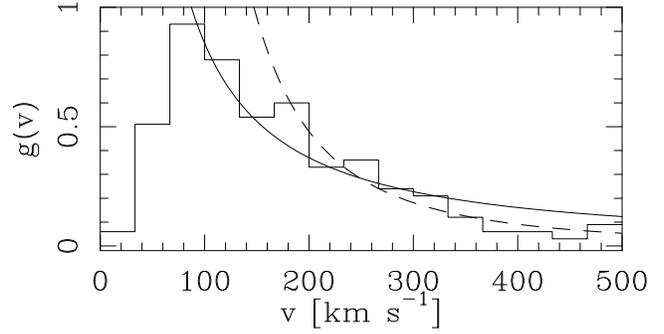}
\caption{The histogram shows the velocity distribution function
  deduced from the observed positions of the tips of the Orion fingers.
  The solid and dashed curved show the power law fits to the observed
  distribution function in the $v\geq 80$ and $v\geq 180$~km~s$^{-1}$
  velocity ranges (respectively).}
\label{fig3}
\end{figure}

The velocity distribution has a steep rise with increasing velocities,
peaks at $v\approx 80$~km~s$^{-1}$, and then has an approximately
monotonic decrease for larger $v$ (see Figure 3). The initial rise
in $g_e(v)$ probably corresponds to incompleteness in the data set,
as a multitude of short, low radial velocity filaments are present
in the CO data (see, e.g., Bally et al, 2017), and are not seen in
H$_2$ images.

Because of this, we carry out power law fits of the form
\begin{equation}
  g(v)=B\left(\frac{\rm 200\,km\,s^{-1}}{v}\right)^\beta\,,
  \label{gfit}
\end{equation}
to the empirical
$g_e(v)$ setting a lower velocity cutoff $v_{min}$ for the fit.
We have tried two values for this cutoff:
\begin{enumerate}
\item $v_{min,1}=80$~km~s$^{-1}$, corresponding to the peak of
  the empirical $g_e(v)$,
\item $v_{min,2}=180$~km~s$^{-1}$, where the decrease of $g_e(v)$
  shows an inflection.
\end{enumerate}
As we can see in Figure 3, the first fit clearly does not
reproduce the decrease of the observed distribution $g_e(v)$
at velocities $>300$~km~s$^{-1}$. For this reason, we favour
the second fit, which does not reproduce the low velocity
behaviour of $g_e(v)$, but this can be understood as being
a result of an incompleteness in the sample of the shorter,
lower velocity filaments and/or of a physical ``soft cutoff''
of the distribution at lower velocities.

With this second fit (see the dashed line of Figure 3), we
obtain $B=(0.48\pm 0.03)$~km$^{-1}$s and $\beta=2.36\pm 0.22$ (see
equation \ref{gfit}). Using
equations (\ref{gv}) and (\ref{gfit}) we see that the exponent
of the corresponding initial radial distribution of the bullets
(see equation \ref{fr}) is $\alpha=(\beta-3)/2=-0.32\pm 0.02$.
Also, comparing equations (\ref{gv}) and (\ref{gfit}), from our
fit we obtain a velocity $v_0=(106\pm 13)$~km~s$^{-1}$ at the outer edge of
the initial disk. We should note that for calculating this
velocity we used the $N=168$ number of bullets in our sample,
which probably has a substantial incompleteness. If we assume that
the disk had a pre-supernova central star of 30~M$_\odot$, this
outer edge velocity implies a disk radius of $\sim 2.4$~AU.

\section{Conclusions}

In a recent paper, Dempsey et al. (2020) modeled the propagation
of orbital debris (protoplanets and/or planets)
dispersed during the breakup of a multiple stellar system. We
study a similar scenario, in which the planets or protoplanets orbit an
embedded massive, pre-supernova star, and travel in approximately
straight trajectories (preserving the modulus of their orbital velocity)
when the supernova ejects most of the stellar mass. This scenario
has no problem in reproducing the high velocities (of up to
$\sim 500$~km~s$^{-1}$) observed in the Orion fingers.

We argue that:
\begin{itemize}
\item the elongated morphology and the division across
  the elongation axis into red and blueshifted regions of
  the ensemble of fingers can be trivially explained as the
  free expansion of an approximately planar, orbital disk
  structure,
\item the velocity distribution of the heads of the Orion fingers
  implies a minimum orbital velocity of $\sim 100$~km~s$^{-1}$,
  and therefore a maximum radius of $\sim 2.4$~AU (for an
  assumed 30 M$_\odot$ mass for the pre-supernova star).
\end{itemize}
From the observed velocity distribution we find that
the initial radial distribution of objects is approximately
$\propto r^\alpha$,
with $\alpha\sim -0.3$. This corresponds to a surface
density of objects with a power law exponent of $\sim -1.3$.

An interesting point is that there is currently no evidence
of the existence of exoplanets or protoplanets orbiting
massive stars or protostars. Possibly the Orion fingers
are evidence in favour of their existence? They
imply the presence of $\sim 100$-200 condensations within
a few AU from the star! A larger size for the initial disk structure
is of course possible if a substantial outwards acceleration
of the condensations (i.e., the protoplanets) occurs when
they interact with the expanding supernova ejection.

We should finally mention that two other systems of ``radiating
fingers'' have been detected in our Galaxy: DR~21 (Zapata et al. 2013,
with only a few bullets) and G5.89 (Zapata et al. 2020, a system
comparable to the Orion fingers). Based on these three detections.
Zapata et al. (2020) conclude that these ``multi-fingered
explosive events'' have a frequency that is similar to the
supernova rate of our galaxy. This result also indicates
that our model deserves further study.

\section*{Acknowledgments}
AR and ARG acknowledge support from the PAPIIT (UNAM) project
IA103121. PRO acknowledges funding from the European Research Council
(ERC) under the European Union's Horizon 2020 research and
innovation program, for the Project ``The Dawn of Organic Chemistry''
(DOC), grant agreement No. 741002. ACR acknowledges support from
a DGAPA-UNAM postdoctoral fellowship.

Data availability:
No new data were generated or analysed in support of this research.

\bsp

\label{lastpage}


\begin{thebibliography}{99}

\bibitem{} Allen, D. A., \& Burton, M. G. 1993, Nature, 363, 54A

\bibitem{} Bally, J., Cunningham, N. J., Burton, M. G. et al.
  2011, ApJ, 727, 113
  
\bibitem{} Bally, J., Ginsburg, A., Arce, H., et al.
  2017, ApJ, 837, 60

\bibitem{} Becklin, E. E, \& Neugebauer, G. 1967, ApJ, 679, L121

\bibitem{} Dempsey, R., Zakamska, N. L., \& Owen, J. E. 2020,
  MNRAS, 495, 1172

\bibitem{} Doi, M., O'Dell, C. R., \& Hartigan, P. 2002, AJ, 124, 445

\bibitem{} G\'omez, L., Rodr\'\i guez, L. F., Loinard, L., et al.
  2005, ApJ, 635, 1166

\bibitem{} G\'omez, L., Rodr\'\i guez, L. F., Loinard, L., et al.
  2008, ApJ, 685, 333

\bibitem{} Kleinmann, D. E., \ Low, F. 1967, ApJ, 149, L1

\bibitem{} Liu, T., Wei, Y.-F., Xue, L. \& Sun, M.-Y. 2021, ApJ, 908, 106L
  
\bibitem{} Moeckl, N., \& Goddi, C. 2012, MNRAS, 419, 1390

\bibitem{} Nissen, H. D., Gustafsson, M., Lemaire, J. L., et al.
  2007, A\&A, 466, 949

\bibitem{} Reipurth, B., Mikkola, S., Connelley, M., et al. 2010,
  ApJL, 725, L56

\bibitem{} Reipurth, B., \& Mikkola, S. 2012, Nature, 492, 221
  
\bibitem{} Reipurth, B., \& Mikkola, S. 2015, AJ, 149, 145
  
\bibitem{} Rivera-Ortiz, P. R., Rodr\'\i guez-Gonz\'alez, A.,
  Cant\'o, J., \& Zapata, L. A. 2021, ApJ, in press (arXiv:2106.01283)

\bibitem{} Rivera-Ortiz, P. R., Rodr\'\i guez-Gonz\'alez, A.,
  Hern\'andez-Mart\'\i nez, L., \& Cant\'o, J. 2019a, ApJ, 874, 38

\bibitem{} Rivera-Ortiz, P. R., Rodr\'\i guez-Gonz\'alez, A.,
  Hern\'andez-Mart\'\i nez, L., et al. 2019b, ApJ, 885, 104

\bibitem{} Rodr\'\i guez, L. F., Poveda, A., Lizano, S., \&
  Allen, C. 2005, ApJ, 627, L65
  
\bibitem{} Zapata, L., Loinard, L., Schmid-Burgk, J., et al.
  2011, ApJ, 726L, 12Z
  
\bibitem{} Zapata, L., Schmid-Burgk, J., P\'erez-Goytia, N.,
  et al. 2013, ApJL, 765, L29

\bibitem{} Zapata, L., Schmid-Burgk, J., Rodr\'\i guez, L. F.,
  et al. 2017, ApJ, 836, 133
  
\bibitem{} Zapata, L., Ho, P. T. P., Fern\'andez-L\'opez, M.,
  et al. 2020,, ApJL. 902, 47


  
  
\end{thebibliography}
\end{document}